\documentclass[aps,superscriptaddress,showkeys,floatfix,amsmath,amssymb,twocolumn,showpacs]{revtex4}
\usepackage{graphicx}
\usepackage{dcolumn}
\usepackage{bm}
\usepackage{epsfig}
\usepackage{xcolor}

\newcommand{\br}{{\bf r}}

\newcommand{\be}{\begin{equation}}
\newcommand{\ee}{\end{equation}}
\newcommand{\bea}{\begin{eqnarray}}
\newcommand{\eea}{\end{eqnarray}}
\begin{document}

\title{A competition between  T=1 and T=0 pairing in $pf$-shell  nuclei with $N=Z$}

\author{H.  Sagawa~$^{\rm 1,2}$,
  Y. Tanimura$~^{\rm 3}$ and K. Hagino~$^{\rm 3}$}
\address{$^{\rm 1}$~Center for Mathematics and Physics, the University of Aizu \\
Aizu-Wakamatsu, Fukushima 965-8580,  Japan\\
$^{\rm 2}$~RIKEN Nishina Center, Wako, Japan \\
$^{\rm 3}$~Department of Physics, Tohoku University, Sendai 980-8578. Japan}


\begin{abstract}
Pairing gain  energies of J=0$^+$, T=1 and J=1$^+$, T=0 states for  the $l=3$ 
and $l=1$ configurations are calculated in the $(1f2p)$ shell model space 
with   T=1 and T=0 pairing interactions,
respectively.   
It is pointed out that two-body matrix element of the spin-triplet 
T=0 pairing is weakened  substantially when the spin-orbit splitting is large in the $1f$ orbits, 
 even the pairing 
strength is much larger than the spin-singlet T=1 pairing interaction. 
 However, the spin-triplet pairing correlations 
may overcome the spin-singlet
  pairing correlations in the $2p$ configuration if the spin-triplet  pairing strength 
 is more than 50\% larger than the 
 spin-singlet pairing.
It is also pointed out in the Hartree-Fock wave functions 
that the mismatching of proton and neutron radial wave functions is at most a few \% level, 
even the Fermi energies are  largely different in the proton and neutron potentials.
 These results imply that the configuration with J=0$^+$, T=1 is very likely in the ground states of odd-odd $pf$ shell nuclei even under the influence of  
 the strong spin-triplet T=0 pairing,   
except at the middle of $pf$ shell in which odd proton and odd neutron may occupy 
 the $2p$ orbits.  These results are consistent with the  observed  spin-parity  J$^{\pi}=0^+$ 
 of all odd-odd $pf$ shell nuclei 
  except $^{70}_{35}$Cu which has J$^{\pi}=1^+$.
\end{abstract}

\maketitle

\section{Introduction}
It has been discussed for a long time the role of the neutron-proton isoscalar spin-triplet 
(T=0,S=1) pairing interaction
 in finite nuclear system \cite{Good1,Macc00} 
 since it is stronger than the isovector spin-singlet (T=1,S=0)  
pairing interaction in the nuclear matter \cite{Cao,BB05}.
  
Nevertheless, the nuclei observed favor the spin-singlet T=1 pairing between identical particles. 
A straightforward answer to this puzzle is that most of stable nuclei have different numbers
  of neutrons and protons, and the T=0 pair  is hardly made  since   the proton and neutron 
occupy the different single-particle  orbits near the Fermi surface.
 Even in nuclei with the equal numbers of protons and neutrons, the $J=1, T=0$ neutron-proton pairing is  not 
a favorable correlation compared  with the $J=0, T=1$ pairing as was seen in the ground state spins of 
odd-odd nuclei in the mass region above $A\geqq 20$ \cite{Macc00}. 
It has been suggested that the nuclear spin-orbit field 
will suppress largely  the spin-triplet pairing than the spin-single pairing \cite{Poves98,Bertsch12,BL10}.  
While so far no clear evidence was found to show  the role of T=0 pairing in the nuclear ground state, the manifestation of the spin-triplet pairing was 
discussed in the high-spin states \cite{Satula,Good2} and also 
in the Gamow-Teller Giant resonences in N=Z nuclei \cite{Bai12}.

We study in this paper the quenching  of two-body matrix element of T=0 pairing interaction in $jj$ coupling 
 scheme in comparison with that of T=1 pairing interaction . Its consequence on the gain energies  is also discussed for the J$^{\pi}$=0$^+$ and the 
  J$^{\pi}$=1$^+$  states in the  $(1f2p)$ shell model configurations by using the HF single-particle 
 wave functions.  The Coulomb interaction is taken into account properly in the HF potential.  
This paper is organized as follows.
We study the two-body matrix elements of  T=0 and T=1  pairing interactions and the overlap of neutron and proton 
HF single-particle states in $(1f2p)$ shell model configuration
  in Section 2.  The competition between the gain energies of T=0 and T=1  pairing interactions
is studied  by diagonalizing the shell model space for $1f$ and $1p$ configurations   in
  Section 3.
 A summary is given in Section 4.

\section{T=0 and T=1 two-body pairing interaction}
We adopt a  separable form of the  pairing interaction in this paper.
The  spin-singlet T=1 pairing interaction is defined as a separable form,
\be
V^{T=1}(\br,\br')=-G^{T=1}\sum_{i,j}P^{(1,0)\dagger}_{i,i}(\br,\br')P_{j,j}^{(1,0)}(\br,\br')
\label{eq:T=1}
\ee
where the pair field  operator reads
\be
P^{(T,S)\dagger}_{a,b}(\br,\br')=
[a^{\dagger}_{a}a^{\dagger}_{b}]^{(T,S)}\psi_a(\br)\psi_{b}(\br').
\ee
 with the single-particle wave function $\psi(\br)$.  
The pairing strength $G^{T=1}$ is fitted to the empirical pairing gaps \cite{BM2,BB05,Poves98} and given by 
\be
G^{T=1}=\frac{24}{A}.
\label{eq:G-T1}
\ee
The absolute value of pairing strength \eqref{eq:G-T1} should not be taken seriously since it depends on the model space adopted.
However, this value might be a reasonable choice for the one major shell model space calculations 
 \cite{BM2,BB05,Poves98}.

The spin-triplet T=0 pairing is also given by a separable  form,
\be
V^{T=0}(\br,\br')=-fG^{T=1}\sum_{i\geqq i',j\geqq j'}P^{(0,1)\dagger}_{i,i'}(\br,\br')
P_{j,j'}^{(0,1)}(\br,\br'))
\label{eq:T=0}
\ee
where $f$ is varied from 1$\sim$2 for  the strength of T=0 pairing interaction.
It should be noticed that, for T=0 pairing, the pair  configurations 
will be   not only in the same orbit with 
  $(l_i=l_{i'},j_i=j_{i'} $ but also in the spin-orbit parter orbits 
with $(l_i=l_{i'},j_i=j_{i'\pm1})$.  
 The two-body matrix element for the T=1 pairing is evaluated to be
\bea
<(j_ij_i)T=1,J=0|V^{(T=1)}|(j_jj_j)T=1,J=0> \nonumber \\
=-\sqrt{(j_i+1/2))(j_j+1/2)}G^{T=1}I_{ij}^2
\label{eq:T=1me}
\eea
where $I_{ij}$ is the overlap integral,  
\bea \label{overlap}
I_{ij}=\int\psi_i(\br)^*\psi_j(\br)d\br
\eea
with  the HF single-particle wave function $\psi(\br)_i$.   
For the T=0 pairing, the two-body matrix element involves the transformation coefficient of $(jj)$ coupling scheme to 
$(ls)$ coupling scheme and reads
\bea
&&<(j_1j_2)T=0,J=1|V^{(T=1)}|(j_1'j_2')T=0,J=1>=\nonumber\\
&& -<[(l_1\frac{1}{2})^{j_1}(l_2\frac{1}{2})^{j_2}]^{J=1}|[(l_1l_2)^{L=0}(\frac{1}{2}\frac{1}{2})^{S=1}]^{J=1}> \nonumber \\
&&<[(l_1'\frac{1}{2})^{j_1'}(l_2'\frac{1}{2})^{j_2'}]^{J=1}|[(l_1'l_2')^{L=0}(\frac{1}{2}\frac{1}{2})^{S=1}]^{J=1}>\nonumber \\
&&\times\frac{\sqrt{2l_1+1}\sqrt{2l_{1}'+1}}{\sqrt{1+\delta_{j_1,j_2}}\sqrt{1+\delta_{j_1',j_2'}}}
 fG^{T=1}(I_{j_1j_1'}I_{j_2j_2'}+I_{j_1j_2'}I_{j_1j_2'})\nonumber \\
\eea
where$ <[(l_1\frac{1}{2})^{j_1}(l_2\frac{1}{2})^{j_2}]^{J=1}|[(l_1l_2)^{L=0}(\frac{1}{2}\frac{1}{2})^{S=1}]^{J=1}>$ is the transformation coefficient  and 
  the overlap integral $I_{ij}$ will involve both the  proton and neuron wave functions.
 The transformation coefficient can be given 
by $9J$ symbol and the explicit form is tabulated  in Table \ref{tab:9j}. 
The square of the  transformation coefficient  is 1/6 and 1/3 for $j_1=j_2$ and $j_1=j_2\pm1$ configurations, respectively,  in the limit
of large angular momentum $l\rightarrow\infty$. These values suggest  large quenchings of the spin-triplet 
pairing correlations and the spin-orbit partner may contribute largely to 
 the spin-triplet pairing matrix.  
While in the small $l$ limit,  $l\rightarrow 0$,  the  coefficient is unity for 
$j=j'=l+1/2$, and  the coefficients are zero for the other  3 configurations.  This suggests that 
the spin-triplet pairing could be large as well as the spin-singlet pairing for the pair configuration in the  
 $s_{1/2}$ orbit, and still 
substantially large  for the configuration in  the  $p_{3/2}$ orbit. 
\begin{table}  \squeezetable \caption{ \label{tab:9j}
The  transformation between $jj$ coupling to $ls$ coupling for the pair wave functions,\\
 $R=<[(l\frac{1}{2})^{j_1}(l\frac{1}{2})^{j_2}]^{J=1}|[(ll)^{L=0}(\frac{1}{2}\frac{1}{2})^{S=1}]^{J=1}>$.
$\Omega$ denotes a value $\Omega=3(2l+1)^2$ .}
\begin{ruledtabular}
\begin{tabular}{cc|c|c|c}
 $j$ &$j'$ & $R$ & $l=1$  & $l=3$  \\\hline
$l+1/2$ &$l+1/2$  &  $\sqrt{\frac{(2l+2)(2l+3)}{2\Omega}}$   & $\frac{1}{3}\sqrt{\frac{10}{3}}$ & $\frac{2\sqrt{3}}{7}$
    \\
$l+1/2$ &$l-1/2$  &  -$\sqrt{\frac{4l(l+1)}{\Omega}}$   & - $\frac{2}{3}\sqrt{\frac{2}{3}}$  & -$\frac{4}{7}$ \\
 $l-1/2$ & $l-1/2$  & - $\sqrt{\frac{2l(2l-1)}{2\Omega}}$   & -$\frac{1}{3}\sqrt{\frac{1}{3}}$ & -$\frac{\sqrt{5}}{7}$ \\
 $l-1/2$  & $l+1/2$ & $\sqrt{\frac{4l(l+1)}{\Omega}}$   & $\frac{2}{3}\sqrt{\frac{2}{3}}$  & $\frac{4}{7}$ \\ 
\end{tabular}
\end{ruledtabular}
\end{table}

 The overlap integral $I_{ij}$ for the neutron-proton pair is performed using HF wave functions 
obtained with a Skyrme interaction SLy4. 
 The single-particle 
 energies of $^{56}$Ni are shown in Fig. 1 for both neutrons and protons.  As is seen in Fig. 1, the  Fermi energies of proton and neutron potentials 
are largely different by about 9 MeV.  Nevertheless the overlap integral of proton and neutron wave functions involved in two-body matrix element have rather 
  similar radial shapes  and the overlap integrals $I_{ij}$ 
are close to 1.0, deviating at most 3\% as is given in Table \ref{tab:overlap}.
 Thus the quenching due to the mismatching of proton and neutron wave functions in the spin-triplet pairing 
 matrix  is rather small compared with that due to the transformation coefficient from ($jj$) to $(LS)$ couplings.    
  Because of this reason, we neglect the mismatching effect of radial wave functions and 
  the overlap integral is taken to be 1 hereafter. The overlap integral of the pair wave functions will 
appear also in the case of short-range  $\delta-$type neutron-proton pairing interaction in which four radial wave functions
are involved in the integral.

\begin{table}
 \squeezetable \caption{ \label{tab:overlap}
 Overlap integrals of proton and neutron HF wave functions in 
  $^{48}$Cr and $^{56}$Ni and $^{64}$Ge.
  The values are given in unit of \%.  
  HF calculations are performed with SLy4 interaction.}
\begin{ruledtabular}
\begin{tabular}{cc|c|c|c}
 $\nu$&$\pi$&    $^{48}$Cr &   $^{56}$Ni  & $^{64}$Ge \\\hline
  1f$_{7/2}$        &  1f$_{7/2} $   & 99.9 & 100. & 99.9   \\
    1f$_{7/2} $       &  1f$_{5/2} $   & 97.7 & 98.9 & 99.1  \\
     1f$_{5/2} $       &  1f$_{7/2}  $  & 99.4 & 99.7 & 99.8   \\
      1f$_{5/2} $       &  1f$_{5/2} $   & 99.6 & 99.8 & 99.9   \\
       2p$_{3/2} $       & 2p$_{3/2}  $  & 99.6 & 99.7 & 99.7   \\
          2p$_{3/2} $       & 2p$_{1/2} $   & 98.2 & 99.1 & 98.9   \\
             2p$_{1/2}  $      & 2p$_{3/2} $   & 99.8 & 99.6 & 99.9   \\
                2p$_{1/2} $       & 2p$_{1/2} $   & 99.1 & 99.6 & 99.6    \\
\end{tabular}
\end{ruledtabular}
\end{table}

\begin{figure}[htp]
\begin{center}
\includegraphics[scale=0.4,clip]{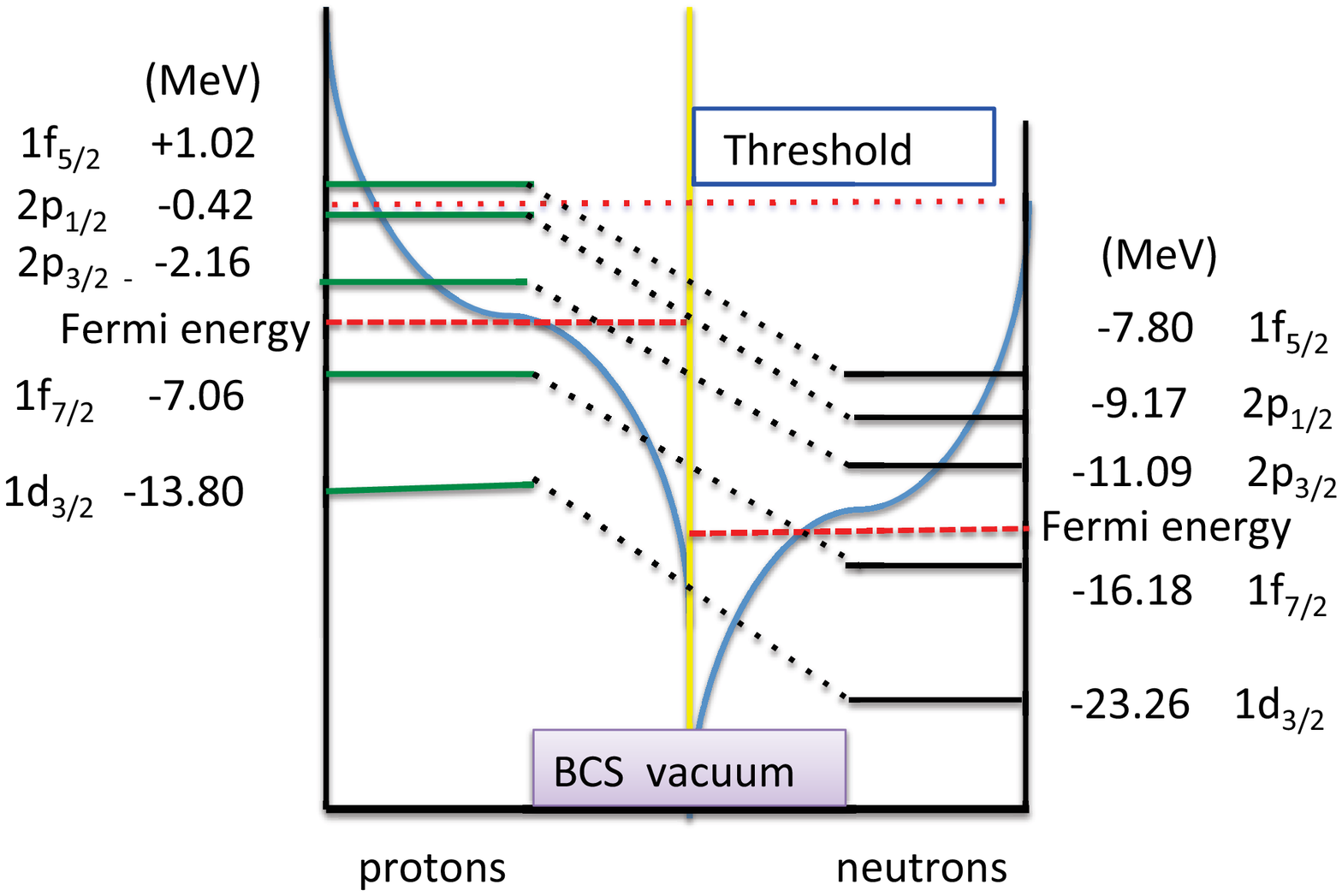}
\caption{HF energies of proton and neutron orbits in  $^{56}$Ni.}
\label{fig:HF}
\end{center}
\end{figure}

\begin{figure}[htp]
\includegraphics[scale=0.4,clip,angle=-90.]{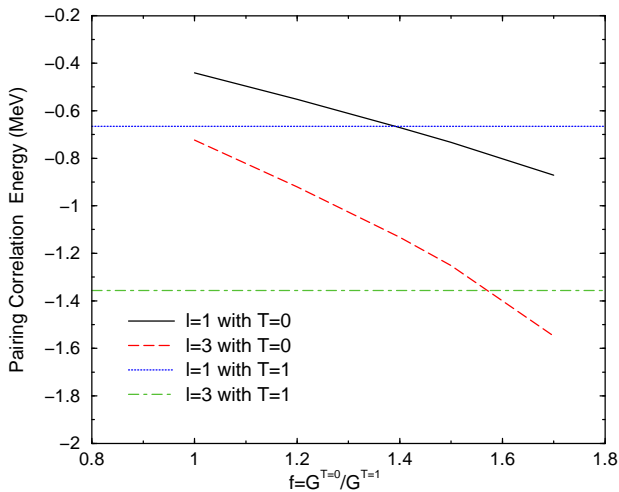}
\caption{Pairing gain  energies of the lowest (J=0$^+$, T=1) and (J=1$^+$, T=0) states for the   $l=3$ 
and $l=1$ configurations.  The spin-singlet T=1 pairing interaction is fixed to be G$^{T=1}$=24/A MeV 
with a mass
A=56, while the spin-triplet  T=0 pairing  is varied with the factor $f=1\sim1.7$ multiplied to the 
 G$^{T=1}$ value.}
\label{fig:BE}
\end{figure}

\section{Pairing gain energy in $pf$ shell configurations}
  In Fig.2, the pairing gain energies for the 
$J^{\pi}=0^+$ state with the isospin T=1 and the $J^{\pi}=1^+$ state  with the isospin 
 T=0 are plotted  as a function of the strength parameter $f$ to the  T=0 pairing interaction taking 
 the $p-$shell ($l=1$) and the $f-$shell ($l=3$) configurations. We diagonalize separately the $p-$ and $f-$shell
 configurations to disentangle the role of the pairing and the spin-orbit interactions in a transparent way.
 For the $l=1$ case, the  ($2p_{3/2})^2$ and ($2p_{1/2})^2$ configurations
 are available for the $J^{\pi}=0^+$ state, while the ($2p_{3/2}2p_{1/2})$ configuration 
is also available for  the $J^{\pi}=1^+$ state.    In a similar way, the 
 ($1f_{7/2}^2)$ and ($1f_{5/2}^2)$ configurations participate to the  $J^{\pi}=0^+$ state in the $l=3$ case,  and  the 
 ($1f_{7/2}1f_{5/2})$ configuration is also involved in  the  $J^{\pi}=1^+$ state.  
  The spin-orbit splitting is parametrized as 
  \be \label{s-o}
  \Delta \varepsilon_{ls} =-V_{ls}({\bf l}\cdot{\bf s}), 
\ee
where the coupling strength is taken to be  \cite{BM1}
 \be \label{s-o-c}
  V_{ls}=\frac{24}{A^{2/3}}.
  \ee
This spin-orbit potential reproduces well the empirical spin-orbit splitting $ \Delta \varepsilon$ =7.0MeV 
between $1f_{7/2}$ and  $1f_{5/2}$ states in $^{41}$Ca \cite{Uozumi}. 
The uncertainly of this strength \eqref{s-o-c}  might be less than 20\% in the
 $sd$ and  $pf$ shell regions even when we adopt  other
 empirical information of the spin-orbit splittings.  
  Using the pairing matrix elements and the spin-orbit splittings, we diagonalize the model space for the $l=1$ and the  $l=3$ configurations, respectively,  
 and  the results are  shown  in Fig. 2.  The lowest energy 
 state with $J^{\pi}$=0$^+$  for the $l=3$ case  gains more binding energy than 
  $J^{\pi}$=1$^+$ state for the strength factor $f< 1.5$.  In the strong T=0 pairing case, 
 $f\geqq 1.6$, the $J^{\pi}$=1$^+$ state obtain more binding energy than the lowest  $J^{\pi}$=0$^+$ state.
   These results are largely due to the quenching of T=0 pairing matrix element
 by  the transformation coefficient from $(jj)$ coupling to $(LS)$ coupling scheme.  This quenching is never happened for the T=1 pairing matrix element  since  the mapping of the two-particle  wave function 
  from $(jj)$ coupling to $(LS)$ coupling
is simply implemented by a factor $\sqrt{j+1/2}$ in Eq. \eqref{eq:T=1me}.  
For the $l=1$ case,   the competition between the  J$^{\pi}$=0$^+$  and the J$^{\pi}$=1$^+$  states
  is shown also in Fig. 2.  Because of the smaller spin-orbit splitting in this case,  the coupling among available configurations are rather strong and   the lowest J$^{\pi}$=1$^+$  state gains more binding energy than the J$^{\pi}$=0$^+$ state in the case $f\geqq 1.4$. These results are consistent with the observed spins of $N=Z$ odd-odd nuclei in the $pf$ shell  where all the ground states have the spin-parity $J^{\pi}=0^+$, except $^{58}_{29}$Cu.
The ground state of $^{58}_{29}$Cu has  $J^{\pi}=1^+$  
  since the odd proton and odd neutron occupy mainly 
 the $2p$ orbits where the spin-orbit splitting is expected to be  much 
smaller than that of $1f$ orbits seen in Fig. 1. 
 
The mass number dependence of the spin-orbit splitting is approximately 
  determined by the Eq. \eqref{s-o}. 
Since the coupling strength and the largest angular momentum in each major shell are proportional to 
$A^{-2/3}$ and $A^{1/3}$ \cite{BM1}, respectively, 
  the spin-orbit splitting of the largest angular momentum states is roughly proportional to $A^{-1/3}$ .
On the other hand, the pairing correlation energy might be proportional to $A^{-1/2}$ as is seen in the
 pairing gap systematics \cite{BB05,BM1b}.  
  Thus, the spin-orbit splitting will decrease slower than the pairing correlation energy as a function of 
the mass number.  As a result, it is expected in medium-heavy  nuclei with $N=Z>30$, that  the spin-orbit splitting 
may hinder more effectively the spin-triplet pairing correlations than lighter nuclei with  $N=Z<30$.  
In reality, 
 the spin-orbit splitting decreases more slowly than the $A^{-1/3}$ dependence;
 6.2 MeV for the $l=1$ states in  $^{16}$O, 5.5 MeV for  the $l=2$ states in $^{40}$Ca, 
  7.0 MeV for the $l=3$ states in $^{56}$Ni and   7.0 MeV for the 
 $l=4$ states in $^{100}$Sn \cite{BM1a,Xavi12}.


It is shown that the shell model matrix elements give the strength factor $f$ 
in Eq. \eqref{eq:T=0} in the range of 1.6-1.7 for both $sd$ shell and
$pf$ shell configurations \cite{BL10,Brown06,Honma04}.  In ref. \cite{Poves98}, the ratio 1.5 is adopted to 
analyze the spin-triplet pairing correlations in the N=Z nuclei in  the shell model calculations.
These adopted values $f$ and the results in Fig. 2  suggests that,  
in the odd-odd 
N=Z nuclei, the $J^{\pi}=1^+$ state  could be a favorite configuration in 
the ground state rather than the $J^{\pi}=0^+$ one especially  
  when the $p_{3/2}$ orbit is the main configuration for the 
valence particles.  However the implementation of spin-triplet pair condensation will not be
guaranteed immediately by the spin of the ground state and nay 
need careful  examination of  many-body wave 
functions obtained by  HF-Bogoliubov or large-scale shell model calculations  
\cite{Gezelis11}. 

\section{SUMMARY}
In summary, we studied the sin-singlet and the spin-triplet pairing correlations in the $pf$ shell model configurations in the nuclei with the same proton and neutron numbers $N=Z$. It is 
   pointed out that the spin-triplet pairing matrix element is largely quenched by the projection of the pair wave function 
from the $(jj)$ coupling scheme to the $(LS)$ coupling scheme,  
On the other hand, there is no quenching in the spin-singlet interaction since   the $J^{\pi}=0^+$ pair in the $(jj)$ 
coupling scheme has the total spin $S=0$ and the projection does not 
 involve any quenching factor.   
  The mismatching of the proton and neutron radial wave functions
 due to the large
difference of the Fermi energies  is also studied by using
  the HF wave functions. While the difference between the proton and neutron 
 Fermi energies is quite large as much as 
9MeV in the $N=Z=28$ nucleus, the overlap integral $I$ between proton and neutron wave function in the spin-triplet pairing matrix
  is rather close to one and the deviation is at most 3\%.
The spin-triplet pairing correlation energy in the $1f$ shell configuration becomes larger than 
the spin-singlet pairing when the scaling factor $f$ of the spin-triplet pairing 
  is larger than 1.6. 
On the other hand, for  the $2p$ configuration, the  spin-triplet pairing correlation becomes dominant   even the factor $f$  is around 1.4.


 
\begin{acknowledgments}
\end{acknowledgments}

\end{document}